\documentclass[aps,amsfonts,prl,nofootinbib,tightenlines]{revtex4}

\usepackage{epsfig}

\input BoxedEPS
\SetRokickiEPSFSpecial  
\HideDisplacementBoxes

\begin{document}

\title{Exact One--Loop Thermal Free Energies of Solitons}

\author{N.~Graham\footnote{e-mail:~graham@physics.ucla.edu}}

\affiliation{Department of Physics and Astronomy \\ University of California
at Los Angeles \\ Los Angeles, CA  90095 \\
{\rm hep-th/0112148 \qquad UCLA/01/TEP/40}}

\begin{abstract}

I show how to compute the exact one--loop thermal correction to the free
energy of a soliton.  The method uses the effective potential as an
auxiliary step to ensure that the soliton is quantized around the
appropriate vacuum.  The exact result is then computed using scattering
theory techniques, and includes all orders in the derivative expansion. 
It can be efficiently combined with a calculation of the exact quantum
correction to yield the full free energy to one loop.  I demonstrate
this technique with explicit computations in $\phi^4$ models,
obtaining the free energy for a kink in $1+1$ dimensions and a domain
wall in $2+1$ dimensions.

{~} 
\end{abstract}

\maketitle

\section{Introduction}

The effective potential is an essential tool in the analysis of the vacuum
states of a quantum field theory.  It provides a computational framework
into which quantum \cite{Coleman} and thermal \cite{Kapusta} corrections
can be efficiently incorporated, allowing for analysis of phenomena
including spontaneous symmetry breaking and restoration.  A key
characteristic of these problems is that they can be analyzed in terms
of constant background fields.  For background fields that are not
constant, the effective potential is still useful as an approximate
calculation, in which the contribution of the higher derivative terms
in the effective action have been ignored.  Once the length scale over
which the field varies is no longer large compared to the Compton
wavelength of the fluctuating fields, however, this approximation is
no longer justified.  In typical soliton problems, these two length
scales are expected to be comparable.  In this case, it is therefore
desirable to replace the derivative expansion with an exact one--loop
computation.

For quantum corrections, such a procedure has been demonstrated in detail
in \cite{method}, which implements ideas originating in
\cite{methodprev}.  In this case, it is essential to precisely
implement the counterterms, which cancel the divergences of the
quantum fluctuations.  To obtain a meaningful finite result, these
counterterms must enforce definite renormalization conditions fixed in
perturbation theory.

Thermal corrections are finite in the ultraviolet, since the contribution
of a mode with energy $\omega$ typically will fall like $e^{-|\omega|/T}$
for large $|\omega|$.  Indeed, they must be finite, since there are no
additional counterterms available beyond those that have already been fixed
at $T=0$.  While the absence of divergences means that the calculation of
thermal corrections doesn't involve the subtleties of renormalization, it
introduces a different problem:  Thermal corrections can modify the
classical vacuum expectation value of the fields.  While quantum
corrections could in principle do so as well, there is a generally
a counterterm available that can cancel this effect.  Standard
renormalization schemes specify that this counterterm be chosen so that the
classical vacuum expectation value of the field is unchanged, which is
represented diagrammatically by the vanishing of the tadpole graph.  Such
counterterms are $T$-independent, however, so they cannot cancel any
further finite shifts in the vacuum caused by thermal corrections.

Since analysis of the vacuum only requires constant fields, the thermal
effective potential is well-equipped to analyze it.  If a soliton exists at
$T=0$, then we expect that for small enough $T$, there should exist
a similar soliton, built around the thermally modified vacuum state.
The original $T=0$ soliton configuration will now have infinitely higher
free energy, since it approaches a value of the field that is not a
minimum of the effective potential.

Once we have constructed the thermally corrected soliton, the thermal
effective potential gives an approximation to its free energy.  But if
we want the exact one--loop thermal free energy, accurate to all
orders in the derivative expansion, we must instead sum the full
contributions from all the small oscillation modes around the thermal
soliton.  At the same time, we must not double-count the approximation
to this quantity already included by shifting the vacuum.  We will
see that the phase shift formalism of \cite{method,methodprev}
provides an efficient mechanism for organizing this calculation.

This approach should be applicable for a wide class of field theory
solitons, allowing for efficient numerical calculation.  Since we do
not use any high--temperature expansions, we can continuously track
the soliton starting from $T=0$.  To illustrate the calculation in a
concrete way, we will choose a simple example, where much of the
computation can be done analytically. We will consider the kink
soliton in $1+1$ dimensions, where the small oscillations potential is
of the exactly solvable reflectionless P\"oschl-Teller form.  We will
also consider the corresponding domain wall in $2+1$ dimensions.
Similar calculations have been considered in this model in
\cite{Carvalho}.  However, that work did not address questions related
to the vacuum shift.  

\section{Effective Potential}

We will consider a $\phi^4$ model in $1+1$ dimensions, with Lagrangian density
\begin{equation}
{\cal L} = \frac{m^2}{2 \lambda} \left(
(\partial_\mu \phi) (\partial^\mu \phi) - \frac{m^2}{4}(\phi^2 - 1)^2 \right)
+ C (\phi^2 - 1)
\end{equation}
where $\phi$ is a real scalar field and $m$ is its classical mass.
Here $C$ is a counterterm, which depends on the cutoff but not on $T$.
At $T=0$, the classical potential is
\begin{equation}
V_c(\phi) = \frac{m^4}{8\lambda} (\phi^2 - 1)^2 \,.
\end{equation}
and the model has a classical soliton solution
\begin{equation}
\phi_{\rm kink}(x) = \tanh \frac{mx}{2}
\end{equation}
which interpolates between the equivalent classical vacua $\phi = \pm 1$.
Once we allow the temperature to be nonzero, we have to take care to
ensure that we continue to consider solitons built around the correct vacua,
which we will find using the quantum and thermal effective potentials.

First, we consider the quantum effective potential, which we compute
at $T=0$.  We use the standard method \cite{Coleman} to compute the
effective potential and fix the counterterms.  We choose a
renormalization scheme in which we hold the location of the minimum of
the effective potential fixed.  This condition is equivalent to
demanding that the tadpole graph vanish, and fixes $C$ uniquely.
Once we have defined the model in this way, the counterterm is fixed
and cannot be changed when we consider $T\neq 0$.  We also add an
overall constant, independent of $\phi$ and $T$, so that the value of the
potential is zero at its minimum.  This calculation yields the quantum
effective potential
\begin{equation}
V_q(\phi) = \frac{1}{8\pi} \left[ U(\phi) \left(1 - \log
\frac{U(\phi)}{m^2}\right) - m^2\right]
\label{quantpot}
\end{equation}
where the small oscillations potential is
\begin{equation}
U(\phi) = \frac{m^2}{2} (3 \phi^2 - 1) = \frac{\lambda}{m^2}
{V_c}''(\phi) \,.
\end{equation}

Following \cite{Kapusta}, the thermal effective potential is given by
\begin{equation}
V_t(\phi) = \frac{T}{\pi} \int_0^\infty \log(1-e^{-\sqrt{q^2 + U(\phi)}/T}) dq
\end{equation}
where $T$ is the temperature.

The total effective potential to one loop is then given by the sum
\begin{equation}
V(\phi) = V_c(\phi) + V_q(\phi) + V_t(\phi) \,.
\label{tot}
\end{equation}
At $T=0$, we have defined the theory so that the minima of this
potential are at $\phi = \pm 1$.  But as $T$ increases from zero, we
cannot prevent the minima from shifting.  For $T$ large enough, the
minimum will move to $\phi=0$ and the symmetry will be restored.  At
that point, the soliton disappears.  For $T$ nonzero but well
below this value, however, we will have a modified soliton.  Our goal
is to compute the difference between its free energy and the free
energy of the trivial vacuum.

We can find the new minimum $\phi_0(T)$ simply by minimizing
eq.~(\ref{tot}), which is easily done numerically.  We must now build
the soliton around this new vacuum.  To accomplish this, we define a
modified classical potential
\begin{equation}
\tilde V_c(\phi) = V_c(\phi) + A(T) (\phi^2 - \phi_0(T)^2) + B(T)
\end{equation}
where we have introduced artificial, finite counterterms $A(T)$ and
$B(T)$.  By choosing
\begin{eqnarray}
A(T) &=& - \frac{V_c'(\phi_0(T))}{2\phi_0(T)}
= -\frac{m^4}{4 \lambda} (\phi_0(T)^2 - 1) \cr
B(T) &=& -V_c(\phi_0(T)) = 
-\frac{m^4}{8 \lambda} (\phi_0(T)^2 - 1)^2
\end{eqnarray}
we force $\tilde V_c(\phi)$ to have its minimum at $\phi_0(T)$, the
same value of $\phi$ as the full effective potential.
Since we will only be interested in the difference between the free
energy of the soliton configuration and the free energy of the trivial
vacuum, we are free to add an overall constant independent of $\phi$,
even one that depends on $T$.  We have used this freedom to  force the
value of the potential at its minimum to be zero.

This modified classical potential is therefore identical to the
zero-temperature potential, except that the artificial counterterm has
shifted the mass of the small oscillations to
\begin{equation}
\tilde m = \sqrt{m^2 + \frac{2\lambda}{m^2} A(T)}
\end{equation}
reflecting the vacuum shift caused by the thermal fluctuations.  The
classical potential is then
\begin{equation}
\tilde V_c(\phi) = \frac{\tilde m^4}{8\lambda\phi_0(T)^2}
(\phi^2 - \phi_0(T)^2)^2
\end{equation}
and the kink solution is
\begin{equation}
\tilde \phi_{\rm kink}(x) = \phi_0(T) \tanh \frac{\tilde m x}{2} \,.
\end{equation}
Since it is built around the correct vacuum, we will be able to use
this soliton for calculations of the quantum and thermal corrections.  
At the end of the calculation, we will use its close relation to the
original potential to subtract the artificial counterterms back out
again, yielding the result in the original model.

\section{Exact Quantum Corrections}

By considering the modified potential $\tilde V_c(\phi)$, we can now find
the quantum corrections computed around the correct vacuum.  The
artificial counterterms $A(T)$ and $B(T)$ summarize the effect of the
thermal contribution in shifting the vacuum.  Later, when we compute the
full thermal correction, we will subtract the contributions of the
artificial counterterms and replace them by the full thermal corrections.
The quantum calculation for $\tilde V_c(\phi)$ now proceeds as a direct
application of the techniques developed in
\cite{method,method1d}. We will outline this calculation, in the
process introducing quantities that will be useful for the thermal
calculation as well.

We consider the small oscillations around the soliton, which are given
by the solutions to the Schr\"odinger equation
\begin{equation}
\left( - \frac{d^2}{dx^2} + \tilde U(\tilde \phi(x))
\right) \psi_k(x) = (k^2 + \tilde m^2) \psi_k(x)
\label{schrodinger}
\end{equation}
where
\begin{equation}
\tilde U(\phi) = \frac{\lambda}{\tilde m^2} \tilde V''(\phi)
\end{equation}
is the modified small oscillations potential.  For the kink, we have
\begin{equation}
\tilde U(\phi_{\rm kink}(x)) = 
\frac{\tilde m^2}{2\phi_0(T)^2} (3 \phi_{\rm kink}(x)^2 - \phi_0(T)^2) = 
\tilde m^2 \left(-\frac{3}{2}{\rm sech}^2 \frac{\tilde m x}{2} + 1 \right) \,.
\label{kinkpot}
\end{equation}

The quantum correction to the energy in the modified theory is given
formally by the sum in the shifts in the zero-point energies of the
small oscillation modes,
\begin{equation}
\tilde{\cal E}_q[\phi(x)] \sim \frac{1}{2} \left(
\sum_{\tilde E_j}|\tilde E_j|-\sum_{\tilde E_j^0}|\tilde E_j^0| \right) 
+\tilde{\cal E}_{ct}[\phi(x)]
\end{equation}
where the modes have energies $\tilde E_j$ in the background $\tilde
\phi(x)$ and energies $\tilde E_j^0$ in the trivial vacuum.
$\tilde{\cal E}_{ct}$ is the contribution of the cutoff-dependent
counterterm.  To express this quantity more precisely, we will work in
the continuum.  Then the sum over modes is replaced by a sum over
bound states plus an integral over scattering states, weighted by the
difference in the density of states between the free and interacting
cases.  For a general soliton background with spherical symmetry,
the spectrum can be decomposed in a partial wave representation into
channels $\ell$ with degeneracy $D_\ell$.  In our one--dimensional
example, these partial waves are just the symmetric and antisymmetric
channels.  In each partial wave, the difference in the density of
states between the free and interacting systems is related to the
scattering phase shift by
\begin{equation}
\rho_\ell(k) - \rho^0_\ell(k) = \frac{1}{\pi} \frac{d\delta_\ell(k)}{dk}
\end{equation}
and there are $n_\ell$ bound states, where
\begin{equation}
\delta_\ell(0) = \pi n_\ell
\end{equation}
which is modified to
\begin{equation}
\delta_+(0) = \pi (n_+ - \frac{1}{2})
\end{equation}
in the case of the symmetric channel in one dimension.  In this case, there
is a ``half-bound'' threshold state in the free background, $\psi=1$.
As the name indicates, such states enter the sum over bound states suppressed
by a factor of $2$ \cite{method1d}. For the case of a charged
particle, we must sum the phase shift for both signs of the energy
corresponding to each value of the momentum $k$.

Following \cite{method1d}, we write the quantum correction as
\begin{equation}
\tilde{\cal E}_q[\phi(x)] = \frac{1}{2} \sum_j \tilde \omega_j
+ \int_0^\infty \frac{dk}{2\pi} \tilde \omega(k) \frac{d}{dk} \tilde \delta(k)
- \frac{\tilde m}{4} + \tilde{\cal E}_{ct}[\phi(x)]
\label{bare}
\end{equation}
where $\tilde \omega(k) = \sqrt{k^2+\tilde m^2}$, the $\tilde
\omega_j$ are the bound state energies, and $\tilde \delta(k) = \tilde
\delta_-(k) + \tilde \delta_+(k) = \frac{1}{2i} \log\det \tilde S(k)$
is the total phase shift in the potential of eq.~(\ref{kinkpot}).  The
$\frac{\tilde m}{4}$ term comes from the energy of the ``half-bound''
state in the free case.  Since the phase shift falls like $1/k$ at
large $k$, the integral in eq.~(\ref{bare}) is logarithmically
divergent, and therefore should be considered as a function of a
regulator, such as the dimension of spacetime \cite{method}.  The
counterterm contribution also depends on the same regulator.  Using
Levinson's theorem we have
\begin{equation}
\tilde{\cal E}_q[\phi(x)] = \frac{1}{2} \sum_j (\tilde\omega_j - \tilde m)
+ \int_0^\infty \frac{dk}{2\pi} (\tilde \omega(k) - \tilde m)
\frac{d}{dk} \tilde \delta(k)
+ \tilde{\cal E}_{ct}[\phi(x)]
\end{equation}
and then we implement the counterterm contribution by expressing it in
terms of the first Born approximation to the phase shift,
\begin{equation}
\tilde\delta_1(k) = -\frac{1}{2k} \int_{-\infty}^\infty 
\left(\tilde U(\phi(x)) - \tilde m^2\right) dx
\end{equation}
giving
\begin{equation}
\tilde{\cal E}_q[\phi(x)] = \frac{1}{2} \sum_j (\tilde \omega_j - \tilde m)
+ \int_0^\infty \frac{dk}{2\pi} (\tilde \omega(k) - \tilde m) \frac{d}{dk}
\left(\tilde\delta(k) - \tilde\delta_1(k) \right) \,.
\label{quantum}
\end{equation}
The Born term depends on the potential only through its integral over
$x$, as it must since that is also how the counterterm depends on the
potential.

In general the phase shifts and bound states can be efficiently
computed numerically.  Eq.~(\ref{kinkpot}), however, is of the exactly
solvable P\"oschl-Teller reflectionless potential form.  We have
\begin{equation}
\tilde \delta(k) = 2\arctan \frac{\tilde m}{k} + 2\arctan \frac{\tilde m}{2k} 
\hbox{~~~~and~~~~} \tilde\delta_1(k) = \frac{3 \tilde m}{k}
\end{equation}
where the bound states are
\begin{equation}
\tilde\omega_0 = 0 \hbox{~~~~and~~~~}
\tilde \omega_1 = \tilde m \frac{\sqrt{3}}{2}
\end{equation}
in addition to a ``half-bound'' threshold state at $\tilde \omega =
\tilde m$.  Plugging these data into eq.~(\ref{quantum}) gives
\begin{equation}
\tilde{\cal E}_q[\phi_{\rm kink}(x)] = 
\tilde m \left(\frac{1}{4\sqrt{3}} - \frac{3}{2\pi}\right)
\end{equation}
for the quantum correction to the modified potential.

\section{Exact Thermal Corrections}

The one--loop thermal correction to the free energy in a background $\phi_0$
is given formally by a sum over the thermal occupation number of the small
oscillation modes,
\begin{eqnarray}
{\cal E}_t[\phi(x)] &\sim& -T  \left(
\sum_{E_j} \log \left(\sum_{n=0}^{\infty} e^{-n |E_j|/T}\right) -
\sum_{E_j^0} \log \left(\sum_{n=0}^{\infty} e^{-n |E_j^0|/T}\right)\right) \cr
&=& T \left (\sum_{E_j} \log (1 - e^{-|E_j|/T}) -
\sum_{E_j^0} \log (1 - e^{-|E_j^0|/T}) \right)
\end{eqnarray}
where we have taken the difference between the interacting and free
backgrounds.  The soliton has a zero mode, corresponding to a
translation rather than an excitation.  It should be omitted from this
sum and handled by collective quantization, representing the motion of
the soliton as a point particle.  We will compute the classical energy
plus the log of the full one--loop determinant with the zero mode
omitted.  This computation corresponds to an isolated soliton at rest.
Ref.~\cite{Coleman} shows how to extend such computations to
a dilute gas of solitons, which requires exponentiating our result and
introducing additional factors of the classical action for each zero mode.

As in the quantum case, we go to the continuum and represent the
density of states in terms of the phase shift.  Decomposing the
scattering into channels $\ell$ with degeneracy $D_\ell$, we have
\begin{eqnarray}
{\cal E}_t[\phi(x)] &=&
T \sum_\ell D_\ell \left(
{\sum_{j}}^\prime \log \left(1 - e^{-|\omega_{j,\ell}|/T} \right) +
\int_0^\infty \frac{dk}{\pi} \log \left(1 - e^{-\omega(k)/T}\right)
\frac{d\delta_\ell(k)}{dk} \right)
\cr
&=& T \sum_\ell D_\ell \left( {\sum_{j}}^\prime \log \left(\frac{1 -
e^{-|\omega_{j,\ell}|/T}}{1 - e^{-m/T}} \right) - \log \left(1 -
e^{-m/T}\right) + \int_0^\infty \frac{dk}{\pi}
\log \left(\frac{1 - e^{-\omega(k)/T}}{1 - e^{-m/T}}\right)
\frac{d\delta_\ell(k)}{dk} \right)
\label{full}
\end{eqnarray}
where we have used Levinson's theorem in the second line.\footnote{For
fermions, where the background field is fixed externally, there is no
vacuum shift.  Thus we can compute the fermion free energy directly as
${\cal E}_t^f[\phi(x)] = -T \sum_\ell D_\ell \left( \sum_{j} \log
\left(1 + e^{-|\omega_{j,\ell}|/T} \right) + \int_0^\infty
\frac{dk}{\pi} \log \left(1 + e^{-\omega(k)/T}\right)
\frac{d\delta_\ell(k)}{dk} \right)$.  This expression gives the
thermal correction to the fermion free energy in a classical
background.  It is similar to the calculation used in \cite{Dunne} for
the thermal correction to the charge.}  The prime indicates that the
zero mode has been omitted, which also accounts for the mismatch with
Levinson's theorem resulting in the second term in the second line.
In our one--dimensional example, this result becomes
\begin{eqnarray}
{\cal E}_t[\phi(x)] = 
T \left(  {\sum_{j}}^\prime \log \left(\frac{1 - e^{-\omega_{j}/T}}{1
- e^{-m/T}}\right) -\log \left(1 - e^{- m/T}\right)
+ \int_0^\infty \frac{dk}{\pi}
\log \left(\frac{1 - e^{-\omega(k)/T}}{1 - e^{-m/T}}\right)
\frac{d\delta(k)}{dk} \right)
\label{thermal}
\end{eqnarray}
which is, as expected, a finite integral.

Next we must modify this calculation to be able to consistently
combine it with our manipulations of the $T=0$ classical
and quantum contributions.  There we introduced an artificial
counterterm as a placeholder for the thermal shift in the vacuum.
First we introduce these same terms into the thermal problem, so that
we write eq.~(\ref{thermal}) in terms of the modified potential $\tilde
U(\tilde \phi_{\rm kink}(x))$ built around the true vacuum
$\phi_0(T)$.  Next we must cancel out all the artificial terms we have
added.  We will accomplish this by imposing a consistency condition:
We know that $\phi_0(T)$ is the local minimum of the full effective
potential.  By construction, it is a minimum of the modified classical
potential, and remained a minimum when the modified quantum
corrections were added, because we arranged to cancel the quantum
tadpole.  Thus if we cancel the modified thermal tadpole with a
counterterm of the same form, we can be sure that we have compensated
for all the previous manipulations.  As with the quantum case,
subtracting the contribution from the first Born expansion implements
this prescription.  It depends only on the background field through
the quantity  $(\tilde U(\tilde \phi_{\rm kink}(x)) - \tilde m^2)$, so it
is indeed of the right form to implement this subtraction.  Thus if we
define
\begin{eqnarray}
\tilde {\cal E}_t[\phi(x)] = T \left(
{\sum_{j}}^\prime \log \left(\frac{1 - e^{-\tilde\omega_{j}/T}}
{1 - e^{-\tilde m/T}}\right) -\log \left(1 - e^{- \tilde m/T}\right)
+ \int_0^\infty \frac{dk}{\pi}
\log \left(\frac{1 - e^{-\tilde\omega(k)/T}}{1 - e^{-\tilde m/T}}\right)
\frac{d}{dk}\left(\tilde \delta(k) - \tilde \delta_1(k)\right) \right)
\label{thermalmod}
\end{eqnarray}
then the full free energy of the true kink system is given by
\begin{equation}
{\cal F}[\phi(x)] = \tilde{\cal E}_c[\phi(x)]
 + \tilde{\cal E}_t[\phi(x)] + \tilde{\cal E}_q[\phi(x)]
\end{equation}
where we have defined the classical contribution
\begin{equation}
\tilde{\cal E}_c[\phi(x)] = \int \tilde V_c(\phi(x)) dx
\end{equation}
which gives
\begin{equation}
\tilde{\cal E}_c[\phi_{\rm kink}(x)] = \frac{\tilde m^3}{3\lambda}
\end{equation}
for the kink.  We have thus made the unique choice such that the free
energy is minimized at the true vacuum $\phi(x) = \phi_0(T)$, and the
value of the energy at this point is zero.

\section{Domain Wall}

It is straightforward to extend this formalism to the case of a domain
wall.  As an example, we will consider the same $\phi^4$ theory in $2+1$
dimensions.  This model was also considered in
\cite{EinhornJones,Cornwall}.  We now have the same kink solution,
where the field is independent of the second spatial coordinate.
Since there are no new divergences, we can continue to use the same
renormalization scheme.  In this case, the quantum effective potential is
\begin{equation}
V_q(\phi) = \frac{1}{12\pi} \left( -U(\phi)\sqrt{U(\phi)} +
\frac{3}{2} m U(\phi) - \frac{1}{2}m^3 \right)
\label{quantpot2}
\end{equation}
and the thermal effective potential is given by
\begin{equation}
V_t(\phi) = \frac{T}{2\pi}
\int_0^\infty q\log(1-e^{-\sqrt{q^2 + U(\phi)}/T}) dq \,.
\end{equation}
The method of \cite{domainwall} gives the full quantum correction to
the energy per unit transverse length as
\begin{equation}
\frac{\tilde{\cal E}_q[\phi(x)]}{L} = -\frac{1}{4\pi} \left(
\frac{1}{2} \sum_j \omega_j^2 \log \frac{\tilde \omega_j^2}{m^2}
+ \int_0^\infty \frac{dk}{2\pi} 
\tilde \omega(k)^2 \log\frac{\tilde \omega(k)^2}{m^2} \frac{d}{dk}
\left(\tilde\delta(k) - \tilde\delta_1(k) \right) \right) \,.
\label{quantum2}
\end{equation}
where the scattering data $\tilde \delta(k)$, $\tilde\delta_1(k)$, and
$\tilde \omega_j$ are unchanged from the case of one space dimension,
since the small oscillation wavefunctions can be separated into
\begin{equation}
\psi_{k,p}(x,y) = \psi_k(x) e^{ipy}
\end{equation}
where $p$ is the momentum in the transverse direction.
Plugging in these data yields
\begin{equation}
\frac{\tilde{\cal E}_q[\phi_{\rm kink}(x)]}{L} = 
\tilde \frac{3m^2}{16 \pi}\left({\rm arccoth}(2) - 2 \right)
\end{equation}
for the quantum correction to the domain wall.
The analog of eq.~(\ref{thermalmod}) is now the thermal free energy
per unit transverse length,
\begin{eqnarray}
\frac{\tilde {\cal E}_t[\phi(x)]}{L} &=& 
T \int_{-\infty}^\infty \frac{dp}{2\pi} 
\left[ {\sum_{j}}^\prime \log \left(\frac{1 - e^{-\tilde\omega_{j}(p)/T}}
{1 - e^{-\tilde m(p)/T}}\right) - \log \left(1 - e^{-\tilde m(p)/T}\right) 
\right. \cr &+& \left.
\int_0^\infty \frac{dk}{\pi}
\log \left(\frac{1 - e^{-\tilde\omega(k,p)/T}}{1 - e^{-\tilde m(p)/T}}\right)
\frac{d}{dk}\left(\tilde \delta(k) - \tilde \delta_1(k)\right) \right]
\label{thermalmod2}
\end{eqnarray}
where $m(p) = \sqrt{m^2+p^2}$, $\omega(k,p)=\sqrt{k^2 + m(p)^2}$ and
$\omega_j(p) = \sqrt{\omega_j^2 + p^2}$.

\section{Examples and Conclusions}

\begin{figure}[hbt]
\centerline{
\BoxedEPSF{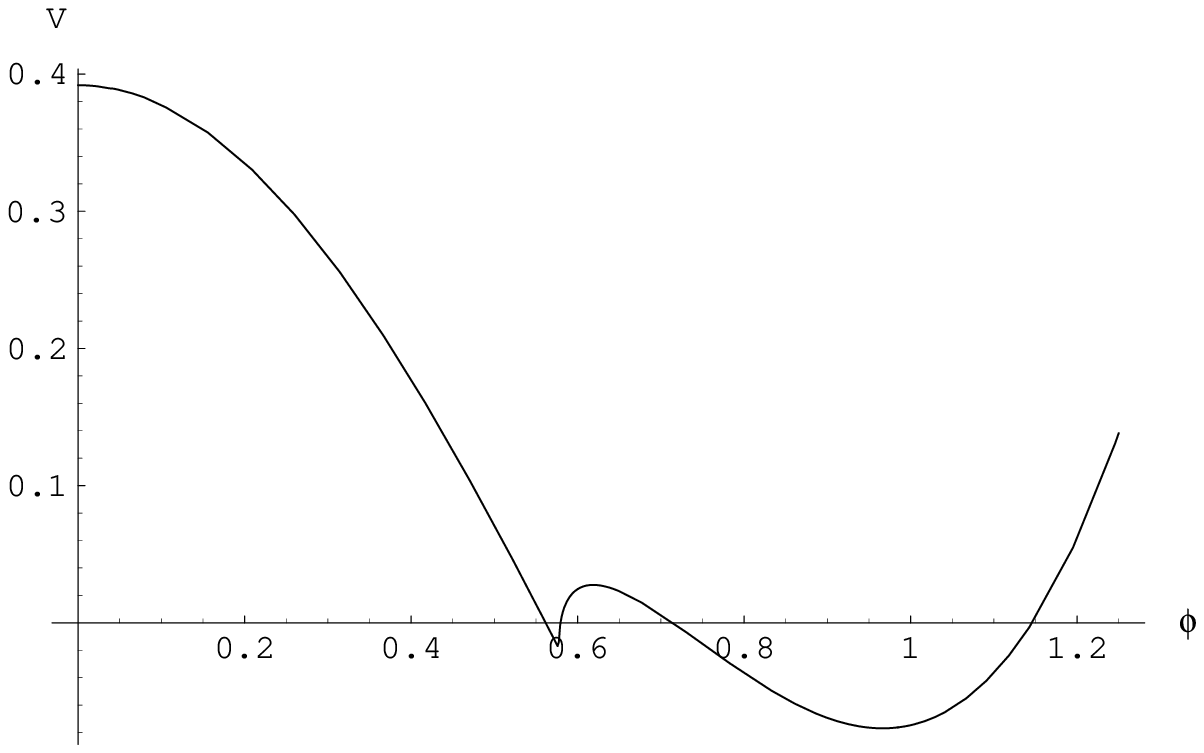 scaled 600}
\hfil
\BoxedEPSF{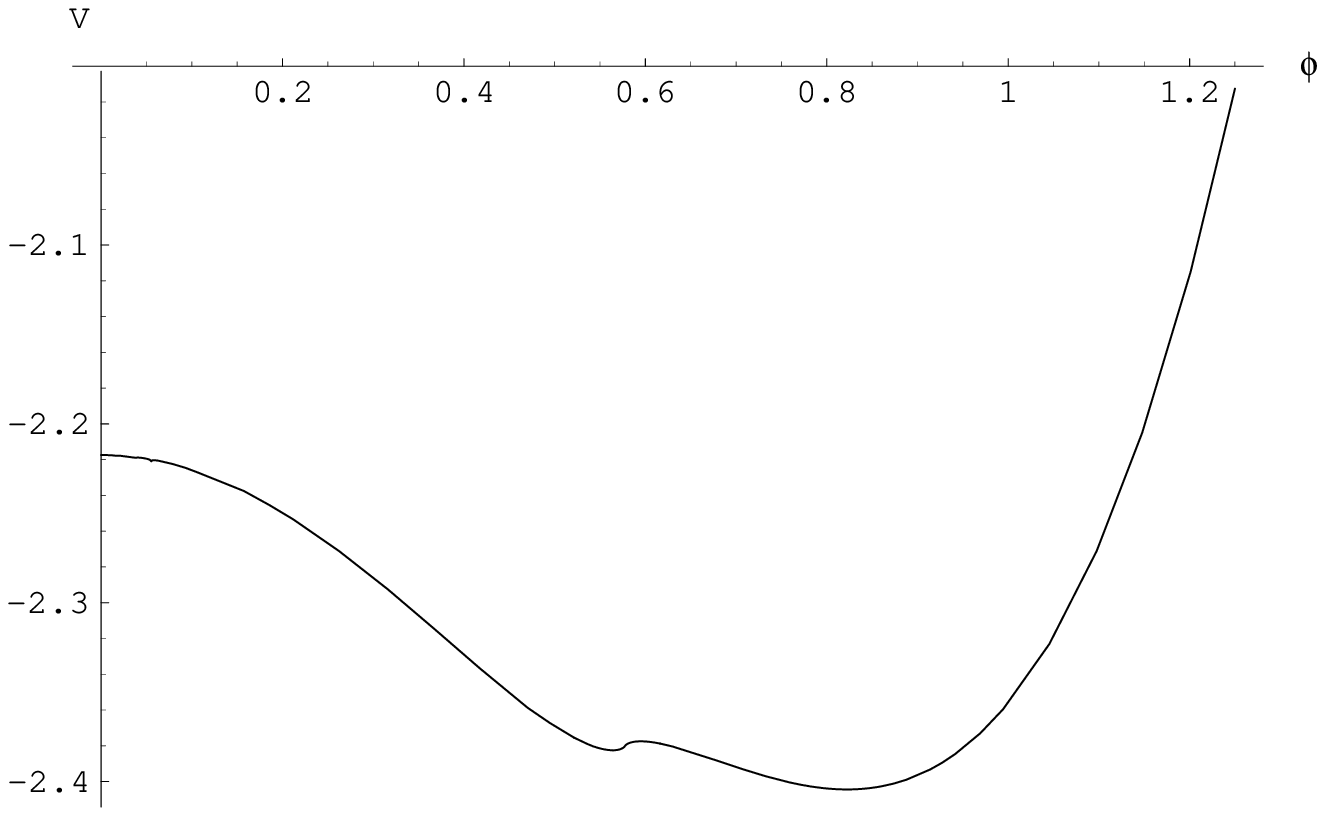 scaled 600}
}
\caption{\sl 
Left:  Real part of the full effective potential $V(\phi)/m^2$ for
the $1+1$ dimensional model with $T=0.7m$ and $\lambda = 0.2 m^2$, as
a function of $\phi$.  
Right:  Real part of the full effective potential $V(\phi)/m^3$ for
the $2+1$ dimensional model with $T=2.4m$ and $\lambda = 0.2 m$, as
a function of $\phi$.
In both cases, the zero--temperature mass $m$ sets the scale of
units.  Away from the minimum, the effective potential can become
complex, which gives a cusp in the plot.}
\label{potplot}
\end{figure}

We illustrate these results with some simple numerical calculations.
Figure~\ref{potplot} shows the full effective potential in $1+1$
dimensions and $2+1$ dimensions, with both plotted as a function of
$\phi$ at fixed $T$.

\begin{figure}[hbt]
\centerline{
\BoxedEPSF{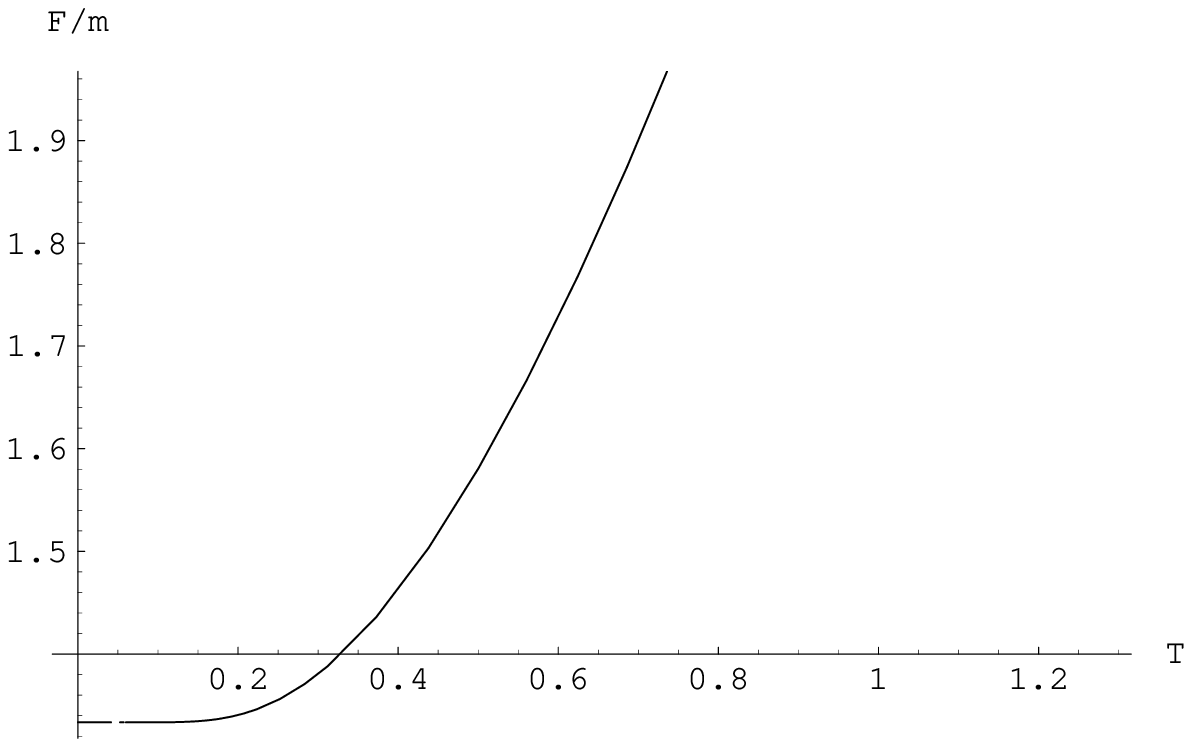 scaled 600}
\hfil
\BoxedEPSF{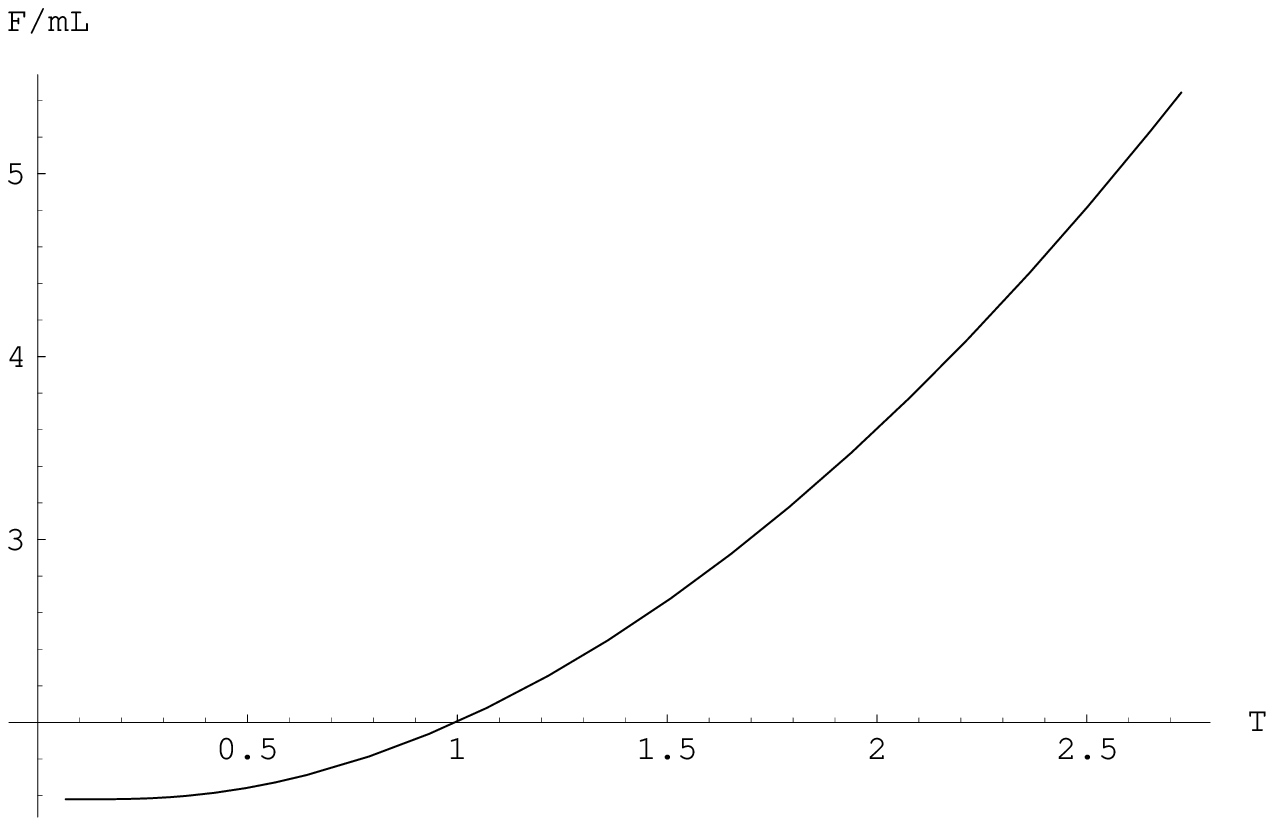 scaled 600}
}
\caption{\sl 
Left:  Full free energy ${\cal F}[\phi(x)]/m$ for the kink soliton
in $1+1$ dimensions as a function of $T/m$, with $\lambda = 0.2 m^2$.
Right:  Full free energy per unit transverse length ${\cal F}[\phi(x)]/mL$
in $2+1$ dimensions as a function of $T/m$ with  $\lambda = 0.2 m$.
In both cases, the zero--temperature mass $m$ sets the scale of units.}
\label{freeplot}
\end{figure}

Figure~\ref{freeplot} shows the full free energy for the soliton and
domain wall.
In both cases, the plot shows the free energy of the extended object
with the free energy of the trivial background subtracted from it,
considered as a function of $T$, with the renormalized
zero--temperature mass and coupling held fixed.  The free
energy cost increases as the temperature increases, as we would expect
since the soliton and domain wall are low--entropy objects.

This technique should be of general use in soliton problems at nonzero
temperature.  Since we have not relied on the high-temperature expansion,
the calculation allows us to track the soliton continuously as we increase
the temperature from zero all the way up to the point where the
soliton melts away.  In models with conserved charges \cite{qball},
one could use this approach to compare the free energy of a soliton
with a state carrying the same charge built on top of the trivial
vacuum.  It could also be used to study phase transitions and find
percolation temperatures in statistical systems.  One might also use
the free energy to estimate rates of soliton formation after phase
transitions.  Finally, we have seen that this approach extends
naturally to strings and domain walls using the techniques developed in
\cite{domainwall}.

\section{Acknowledgments}

I would like to thank  C.~A.~de Carvalho, J.~M.~Cornwall, G.~Dunne, and
A.~Kusenko for helpful discussions.  This work is supported by the
U.S.~Department of Energy (D.O.E.) under cooperative research
agreement \#DE-FG03-91ER40662.


\begin{thebibliography}{99}


\bibitem{Coleman}
S.~Coleman, {\sl Aspects of Symmetry} (Cambridge University Press, 1985).

\bibitem{Kapusta}
L.~Dolan and R.~Jackiw, Phys.~Rev~{\bf D9} (1974) 3320;
J.~Kapusta, {\sl Finite-temperature field theory }
(Cambridge University Press, 1989).

\bibitem{method}
E.~Farhi, N.~Graham, P.~Haagensen and R.~L.~Jaffe, Phys.~Lett.~{\bf B427}
(1998) 334; E.~Farhi, N.~Graham, R.~L.~Jaffe, and H.~Weigel,
Nucl.~Phys.~{\bf B585} (2000) 443; Nucl. Phys. {\bf B595} (2001) 536.

\bibitem{methodprev}
J.~Schwinger, Phys.~Rev.~{\bf 94} (1954) 1362;
J.~Baacke, Z.~Phys.~{\bf C53} (1992) 402.

\bibitem{Carvalho}
C.~A.~de Carvalho, arXiv:hep-ph/0110103.

\bibitem{method1d}
N.~Graham and R.~L.~Jaffe, Phys.~Lett.~{\bf B435} (1998) 145;
N.~Graham and R.L.~Jaffe, Nucl. Phys.~{\bf B544} (1999) 432;
N.~Graham and R.L.~Jaffe, Nucl. Phys.~{\bf B549} (1999) 516.

\bibitem{Dunne}
G.~V.~Dunne and K.~Rao, Phys.~Rev.~{\bf D64}, 025003 (2001);
A.~J.~Niemi and G.~W.~Semenoff, Phys.~Lett.~{\bf B135} (1984) 121.

\bibitem{Cornwall}
C.~A.~de Carvalho, J.~M.~Cornwall and A.~J.~da Silva,
Phys.~Rev.~{\bf D64} (2001) 025021.

\bibitem{EinhornJones}
M.~B.~Einhorn and D.~R.~T.~Jones,
Nucl.~Phys.~{\bf B398} (1993) 611.

\bibitem{domainwall}
N.~Graham, R.~L.~Jaffe, M.~Quandt, and H.~Weigel,
Phys. Rev. Lett.  {\bf 87} (2001) 131601.

\bibitem{qball}
A.~Kusenko, Phys.~Lett.~{\bf B404} (1997) 285;
M.~Laine and M.~E.~Shaposhnikov, Nucl.~Phys.~{\bf B532} (1998) 376;
N.~Graham, Phys.~Lett.~{\bf B513} (2001) 112.

\end{thebibliography}
\end{document}